\documentclass[conference,a4paper]{IEEEtran}
\IEEEoverridecommandlockouts

\usepackage{cite}
\usepackage{amsmath,amssymb,amsfonts}
\usepackage{lipsum}
\usepackage{mathtools}
\usepackage{cuted}
\usepackage{algorithm}
\usepackage{algorithmic}
\usepackage{graphicx}
\usepackage{textcomp}
\usepackage{xcolor}
\usepackage{multirow}
\def\BibTeX{{\rm B\kern-.05em{\sc i\kern-.025em b}\kern-.08em
    T\kern-.1667em\lower.7ex\hbox{E}\kern-.125emX}}
\usepackage{etoolbox}
\include{misc/user_colors}
\makeatletter
\patchcmd{\@makecaption}
  {\scshape}
  {}
  {}
  {}
\makeatother

\newcommand{\bup}{\mathbf{b}}
\newcommand{\qbup}{\mathbf{\hat{b}}}

\newcommand{\bslen}[1]{|#1|}
\newcommand{\bbl}{\bslen{bb}}
\newcommand{\nbl}{\bslen{nb}}

\newcommand{\cheng}{\textbf{ch}}
\newcommand{\bmshj}{\textbf{bm}}

\begin{document}

\title{Improving The Reconstruction Quality by Overfitted Decoder Bias in Neural Image Compression
}

\author{\IEEEauthorblockN{Oussama Jourairi}
\IEEEauthorblockA{\textit{InterDigital, Inc.}\\
Rennes, France \\
}
\and
\IEEEauthorblockN{Muhammet Balcilar}
\IEEEauthorblockA{\textit{InterDigital, Inc.}\\
Rennes, France \\
}
\and
\IEEEauthorblockN{Anne Lambert}
\IEEEauthorblockA{\textit{InterDigital, Inc.}\\
Rennes, France \\
}
\and
\IEEEauthorblockN{François Schnitzler}
\IEEEauthorblockA{\textit{InterDigital, Inc.}\\
Rennes, France \\
}
}

\IEEEoverridecommandlockouts
\IEEEpubid{\makebox[\columnwidth]{IEEE copyright:~\copyright2022 IEEE, all rights reserved\hfill} \hspace{\columnsep}\makebox[\columnwidth]{ }}

\maketitle

\begin{abstract}
End-to-end trainable models have reached the performance of traditional handcrafted compression techniques on videos and images. Since the parameters of these models are learned over large training sets, they are not optimal for any given image to be compressed. 
 In this paper, we propose an instance-based fine-tuning of a subset of decoder's bias to improve the reconstruction quality in exchange for extra encoding time and minor additional signaling cost. The proposed method is applicable to any end-to-end compression methods, improving the state-of-the-art neural image compression BD-rate by $3-5\%$.  
\end{abstract}


\section{Introduction}
\label{sec:intro}
Image and video compression are an important part of our everyday life. These technologies have been refined over decades by experts. Nowadays, compression algorithms, such as those developed by MPEG, consist in fine-tuned handcrafted techniques.
Recently, deep learning models have been used to develop end-to-end trainable compression algorithms.
State of the art neural architectures now compete with traditional compression methods (H.266/VVC \cite{vvc}) even in terms of peak signal-to-noise ratio (PSNR) for single image compression \cite{gao2021neural}.

One of the main research directions for end-to-end compression focuses on  
Rate-Distortion Autoencoder \cite{habibian2019video}, a particular type of Variational Autoencoder (VAE) models \cite{kingma2013auto}. 
Optimizing such a model amounts to minimizing the mean square error (MSE) of the decompressed image and the bitlength of encoded latent values, estimated by their entropy w.r.t their priors \cite{balle2018variational}. In practice, these latents are first quantized and then typically encoded by an entropy encoder such as range or arithmetic coding \cite{1056282}. These encoders exploit the prior distributions over the encoded values (here, the latents) to achieve close to optimal compression rates.
The priors are also trainable and can themselves have hyperpriors \cite{theis2017lossy,balle2016end,balle2018variational,minen_joint,minnen2020channel,cheng2020image,xie2021enhanced}.

As usual with deep learning, these models are typically trained on large datasets and fixed whereas traditional encoders can adapt to a particular image by for example optimizing the quadtree decomposition. So, any resulting neural model is likely to be suboptimal for any single image, a problem called the amortization gap \cite{cremer2018inference}. In the compression context, this can be leveraged to improve the rate-distortion trade-off, for example by fine-tuning the encoder or the latent codes \cite{LuCZCOXG20,Campos_2019_CVPR_Workshops,NEURIPS2020_066f182b,Guo_2020_CVPR_Workshops}. These approaches improve distortion without degrading the rate. Another class of methods fine-tunes the decoder and the entropy model, improving distortion further but degrading the rate, as modified parameters must be transmitted as well \cite{9287069,Aytekin_2018_CVPR_Workshops,van2021overfitting}. Because of this added cost, these approaches have not been applicable to single image compression but only to set of images \cite{van2021overfitting} or video \cite{van2021instance}, where the rate increase is amortized over many images. Another solution is to select one set of parameter values out of predefined sets \cite{9287069}. This decreases encoding time and signaling cost but has again limited gain compared to strong baselines. 

In this paper, we achieve decoder fine-tuning to improve the reconstruction quality for single image compression which was found infeasible in the literature so far. This is made possible thanks to our three contributions: 1) selection of subset of parameters to be fine-tuned, 2) learning the quantization parameter of updates jointly and 3) using a new loss function based on interpolation of the baseline model's performances. 
In our experiment, we show  $3-5\%$ BD-rate gain for any given baseline end-to-end image compression model in exchange for extra encoding complexity.

\section{Neural Image Compression}
\label{sec:endtoend}
An input image to be compressed,  $\mathbf{x} \in \mathbb{R}^{n\times n \times 3}$ is first processed by a deep encoder $\mathbf{y}=g_a(\mathbf{x};\mathbf{\phi})$. $\mathbf{y} \in \mathbb{R}^{m\times m \times o}$  is called the latent and is smaller than $\mathbf{x}$. This latent is converted into a bitstream by going through a quantizer, $\mathbf{\hat{y}}=\mathbf{Q}(\mathbf{y})$, and then through an entropy coder exploiting a prior $p_{f}(\mathbf{\hat{y}}|\Psi)$ in \cite{balle2016end}.
$p_f$ can also depend on some side information $\mathbf{z}=h_a(\mathbf{y})  \in \mathbb{R}^{k\times k \times f}$ to better model spatial dependencies. $h_a$, another neural network, is also trained. We denote by $\mathbf{\hat{z}}=\mathbf{Q}(\mathbf{z})$ the quantization of $\mathbf{z}$. 
Both $\mathbf{\hat{y}}$ are $\mathbf{\hat{z}}$ are encoded and the encoders respectively use the hyperpriors $p_{h}(\mathbf{\hat{y}}|\mathbf{\hat{z}};\Theta)$, and  $p_{f}(\mathbf{\hat{z}}|\Psi)$.  
The latent can be processed by a deep decoder $\mathbf{\hat{x}}=g_s(\mathbf{\hat{y}};\mathbf{\theta})$ to obtain the decompressed image $ \mathbf{\hat{x}}$.
The parameters $\mathbf{\phi}, \mathbf{\theta}, \Psi, \Theta $ are trained using the following rate-distortion loss:
\begin{equation}
\small
   \label{eq:balle2}
   \mathcal{L}=\mathop{\mathbb{E}}_{\substack{\mathbf{x}\sim p_x \\ \epsilon \sim U}}\left[-\log(p_{h}(\mathbf{\hat{y}}|\mathbf{\hat{z}},\Theta)) -\log(p_{f}(\mathbf{\hat{z}}|\Psi)) + \lambda d(\mathbf{x},\mathbf{\hat{x}})\right],
\end{equation}
where, $d(.,.)$ denotes a distortion loss such as MSE and $\lambda$ controls the trade-off between compression ratio and quality. Note that during training $\mathbf{Q}(.)$ is relaxed into $\mathbf{Q}(x)=x+\epsilon$, $\epsilon \sim \mathcal{U}(-0.5,0.5)$. 

Typically, $p_{f}(\mathbf{\hat{z}}|\Psi)$ is factorized in $f$ independent slices of size $k \times k$. Each slice has its own trainable cumulative distribution function (cdf): $\bar{p}_{\Psi}^{(c)}(.).  
~c=1\dots f$, 
From the cdf and for any  value of $x$, the probability mass function (pmf) is derived 
by $\hat{p}_{\Psi}^{(c)}(x)=\bar{p}_{\Psi}^{(c)}(x+0.5)-\bar{p}_{\Psi}^{(c)}(x-0.5)$.
Hence,
\begin{equation}
   \label{eq:factorent}
   p_{f}(\mathbf{\hat{z}}| \Psi)=\prod_{c=1}^{f} \prod_{i,j=1}^{k,k} \hat{p}_{\Psi}^{(c)}({\mathbf{\hat{z}}_{i,j,c}})\enspace.
\end{equation}

Similarly, in $p_{h}(\mathbf{\hat{y}}|\mathbf{\hat{z}};\Theta)$, the distribution of each latent value is assumed to follow a $1d$ Gaussian distribution and its pmf is $\hat{\mathcal{N}}(x;\mu,\sigma)=\bar{\mathcal{N}}(x+0.5;\mu,\sigma)-\bar{\mathcal{N}}(x-0.5;\mu,\sigma)$ and therefore during training $p_{h}(\mathbf{\hat{y}}|\mathbf{\hat{z}},\Theta) = \prod_i \hat{\mathcal{N}}(\mathbf{\hat{y}}_i;\mu_i,\sigma_i)$ 
where $\mu,\sigma$ are computed by a neural network $h_s$ parameterized by $\Theta$. Various architectures have been considered, for example $h_s(\mathbf{\hat{z}};\Theta)$  \cite{balle2018variational} 
or  
 autoregressive predictions
$\mu_i,\sigma_i=h_s(\mathbf{\hat{z},\mathbf{\hat{y}}_{<i}};\Theta)$  \cite{minen_joint,cheng2020image,xie2021enhanced}.

A trained $g_s$ performs well on average but is likely to be suboptimal for any single image. It is possible to improve the rate-distortion trade-off for a video by retraining $\theta$ specifically for this video and by transmitting quantized weight updates $\mathbf{\hat{\delta}}$ for the decoder in addition to the quantized latents \cite{van2021overfitting,van2021instance}. The loss used for fine tuning can for example be:
\begin{equation}
\small
   \label{eq:ft}
   \mathcal{L}= -\log(p(\mathbf{\hat{\delta}}|\Phi)) + \gamma d(\mathbf{x},g_s(\mathbf{\hat{y}};\mathbf{\theta}+\mathbf{\hat{\delta}})) \enspace,
\end{equation}
where $p$ is the probability of quantized weight updates, parameterized by $\Phi$.
\section{Method}
\label{sec:method}
Detailed below, our instance based post-training method jointly optimizes a predefined subset of convolution bias in a decoder, using a novel loss guarantying better RD performance, and the parameters of the bias updates quantization. 

\subsection{Selection of Parameters to be Overfitted}
\label{sec:selection}
Any parameter in $\mathbf{\phi},\mathbf{\theta},\Phi,\mathbf{\Theta},\Psi$ causes some amortization gap and can be overfitted to the given instance to close this gap. Since the decoder ($g_s(.;\mathbf{\theta})$) is the most important block to improve reconstruction, we focus on this part. Even limited to $\mathbf{\theta}$, a state-of-the-art neural model such as \cite{cheng2020image} has millions of parameters (generally more than the number of pixels in a given image to reconstruct any image easily). This makes it infeasible to signal updates of these parameters for single image compression. Therefore, we propose to overfit only some parameters. 
Since the bias terms in the convolution layers are arguably the most important (more sensitive to a given instance) and less numerous than convolution weights, we prefer to overfit the bias terms only.
Furthermore, it is known that the first layers of networks mostly relate to the more general features and last layers to the details. Thus, to improve reconstruction details, we further limit overfitting to the bias terms of the last $l$ convolution layers. Overfitting the biases for sets of images was also investigated in \cite{9287069}. Formally, we divide the decoder parameter set into two parts: $\mathbf{\theta}=[\mathbf{\theta}_g,\beta_g]$, where $\beta_g \in \mathbb{R}^{u}$ are the bias terms in the last $l$ convolution layers and $\mathbf{\theta}_g \in \mathbb{R}^{v}$ the other decoder parameters where $v >> u$. In our approach, the decoder function can be seen as $g_s(.;\mathbf{\theta}_g,\mathbf{\beta}_g+\bup)$, where $\bup$ is the vector of overfitted weight updates. The quality of the selections of subset of biases is experimentally validated in Section \ref{sec:results}.

\subsection{Interpolation Based Loss}
\label{sec:loss}
Neural image compression models are trained with different $\lambda$s (Eq. \eqref{eq:balle2}), yielding different points of the RD curve. For each set of parameters, the bit-rate can be measured by $\mathbf{R}=-\log(p_{h}(\mathbf{\hat{y}}|\mathbf{\hat{z}},\Theta)) -\log(p_{f}(\mathbf{\hat{z}}|\Psi))$ in terms of total bit length and the reconstruction quality by $\mathbf{D}(\mathbf{x},\mathbf{\hat{x}})$ in terms of peak
signal-to-noise ratio (PSNR). 
To compare two models,  BD-rate\cite{bjontegaard2001calculation} is often used. It is derived from the RD curve by interpolating $\mathbf{D}$ for any continuous $\mathbf{R}$. Since the interpolation is done by higher order polynomial fitting and is a monotonically increasing function, this  $\mathbf{D}(\mathbf{R})$ function is invertible. Thus, we can obtain a $\mathbf{R}(\mathbf{D})$ function.   

The relative bit-saving of the new model can be defined as the ratio of the bitlength of the new model ($\nbl$) and of the baseline model ($\bbl$) under the same distortion by $1-\frac{\nbl}{\bbl}$. It motivates us to minimize  $\frac{\nbl}{\bbl}$ to ensure the best bit-savings. During the optimization, the distortion of our model changes and  these distortion values have probably not been sampled for the baseline model RD curve. Thus, we cannot directly obtain the rate of the baseline models for these values. However, we can use the baseline model $\mathbf{R}(\mathbf{D})$ function to approximate the bitlength for any given distortion. Thus, our method needs the baseline model $\mathbf{R}(\mathbf{D})$ function in addition to a trained baseline model for one single quality (starting point for overfitting). We then minimize the following loss function:     
\begin{equation}
\small
   \label{eq:newloss}
   \mathcal{L}=\mathop{\mathbb{E}}_{\epsilon \sim U}
   \left[
   \frac{\mathbf{R}\left(\mathbf{d}(\mathbf{x},g_s(\mathbf{\hat{y}};\mathbf{\theta}_g,\beta_g))\right)- \log(p(\mathbf{Q}(\bup;q)))+C}
   {\mathbf{R}\left(\mathbf{d}(\mathbf{x},g_s(\mathbf{\hat{y}};\mathbf{\theta}_g,\beta_g+\mathbf{Q}^{-1}\circ\mathbf{Q}(\bup;q)))\right)} 
   \right],
\end{equation}

where the trainable parameters are bias updates $\bup \in \mathbb{R}^{u}$ and quantization scale parameter $q \in \mathbb{R}$, which is defined in the next section. $\mathbf{Q}(.)$ and $\mathbf{Q}^{-1}(.)$ respectively refer to quantization and dequantization steps, which are controlled by $q$. The probability of quantized bias update symbols is denoted by $p(.)$. 
The denominator is the bitlength of the baseline model (estimated by $\mathbf{R}(\mathbf{D})$) for the distortion achieved by our approach. 
The numerator is the expected bitlength of the proposed model. It is always larger than the (fixed) baseline bitlength ($\mathbf{R}\left(\mathbf{d}(\mathbf{x},g_s(\mathbf{\hat{y}};\mathbf{\theta}_g,\beta_g))\right)$) as it also includes the bitlength of bias updates ($-\log(p(\mathbf{Q}(\bup;q))$) and the bitlength of extra information ($C \in \mathbb{R}$ is fixed). Even though our bitlength is longer than the baseline model, the resulting rate is smaller than what the baseline model could achieve for that distortion.

\subsection{Quantization and Entropy Coding of Updates}
\label{sec:quant}
We use uniform scalar quantization over scaled bias updates in test time. We implement quantization by rounding the scaled inputs to the nearest integer value by $\mathbf{Q}(\bup;q)=round(\bup.q)$. Since the $q$ value is learned jointly, the model may adjust the quantization resolution. Dequantization cancels the scaling, thus at test time $\mathbf{Q}^{-1}\circ\mathbf{Q}(\bup;q)=round(\bup.q)/q$. However, since the rounding operator has non-informative gradients, we cannot use it for training. Continuous relaxation of rounding operator with additive uniform noise is used at training time as in \cite{balle2018variational}. Thus, for training,  quantization and dequantization are as follows;
\begin{equation}
   \label{eq:quant}
\mathbf{Q}(\bup;q)=\bup.q + \mathbf{\epsilon},~~~~~ \mathbf{Q}^{-1}\circ\mathbf{Q}(\bup;q)=\bup + \mathbf{\epsilon}/q,
\end{equation}

\noindent where $\mathbf{\epsilon} \in \mathbb{R}^u$ is iid uniform noise where $\mathbf{\epsilon}_i \sim U(-0.5,0.5)$. Since $q$ is learned, it is included in the bitstream.

The bias updates follow a Gaussian distribution. Since we quantize the scaled updates to the nearest integer value, the bin width of the quantization is 1. Thus, the expected probability of the given scaled and quantized update vector $\qbup$ can be calculated at training as follows;

\begin{equation}
   \label{eq:prob}
p({\qbup})=\prod_i \int_{{\qbup}_i-0.5}^{{\qbup}_i+0.5} \mathcal{N}(x;\mu,\sigma).dx
\end{equation}

\noindent where $\mathcal{N}(.;\mu,\sigma)$ is the probability density function of a Gaussian distribution parameterized by $\mu,\sigma$ which are the mean and standard deviation of vector ${\qbup}$ (the closed form solution to fit a Gaussian probability model on ${\qbup}$). At test time, to compress $\qbup$ with entropy coding, we fit the truncated Gaussian distribution on $\qbup$. Its support is defined from minimum symbol $s_{min}$ to maximum symbol $s_{max}$.
The parameters of the fitted truncated Gaussian are also included in the bitstream: $\mu,\sigma$, both using 16-bits and $s_{min},s_{max}$ using 8-bits for each parameter. With the 16-bits used for $q$, the bitlength of the extra information is 64 ($C$ in loss function in Eq. \eqref{eq:newloss}).

\subsection{Encoding and Decoding}
\label{sec:encdec}

The encoding and decoding flows of the proposed model are described by Algorithms \ref{alg:encode} and \ref{alg:decode} respectively. As it is shown, our method uses an extra bitstream $eb$ in addition to the main latent bitstream $mb$ and side latent bitstream $sb$ in the baseline model. Their empirical lengths are $|mb|,|sb|$ and $|eb|$. Our results were computed by considering the fact that baseline model bitlength is $|mb|+|sb|$ but ours is $|mb|+|sb|+|eb|$.   
\begin{algorithm}[tb]
  \caption{Proposed Encoding}
  \label{alg:encode}
  \footnotesize
\begin{algorithmic}
  \STATE {\bfseries Input:} Image $\mathbf{x}$, Trained model components $g_a(.;\mathbf{\phi})$, $h_a(.;\Phi)$, $g_s(.;\mathbf{\theta}_g,\beta_g)$, $p_{f}(.|\Psi)$, $p_{h}(.|.;\Theta)$, $\mathbf{R}(\mathbf{D})$ function, fixed cost $C$.
  \STATE {\bfseries Output:} Main latent bitstream $mb$, side latent bitstream $sb$ and extra bitstream $eb$.
 \vspace{0.05cm}
 \hrule
 \vspace{0.05cm}
  \STATE $\mathbf{y}=g_a(\mathbf{x};\mathbf{\phi})$; $\mathbf{z}=h_a(\mathbf{y};\mathbf{\Phi})$; $\mathbf{\hat{z}}=\mathbf{Q}(\mathbf{z})$; $\mathbf{\hat{y}}=\mathbf{Q}(\mathbf{y})$.
  \STATE $sb.write(\mathbf{\hat{z}},p_{f}(.|\Psi))$ // encode side latent w.r.t factorized entropy
  \STATE $mb.write(\mathbf{\hat{y}},p_{h}(.|\mathbf{\hat{z}};\Theta))$ // encode main latent w.r.t hyperprior entropy
  \STATE Initialize $\bup$ and $q$.
  \FOR{$epoch=1$ {\bfseries to} $maxiter$} 
  \STATE // Train time
  \STATE $\epsilon \sim U(-0.5,0.5)$ // generate random uniform noise
  \STATE $\qbup=\bup.q+\epsilon$
  \STATE $psnr=\mathbf{d}(\mathbf{x},g_s(\mathbf{\hat{y}};\mathbf{\theta}_g,\beta_g+\qbup/q))$
  \STATE $rate=-log(p(\qbup))+C$
  \STATE $\mathcal{L}= \frac{|mb|+|sb|+rate}{\mathbf{R}(psnr)}$ 
  \STATE Update $\bup$ and $q$ w.r.t $d\mathcal{L}/db$ and $d\mathcal{L}/dq$
  \STATE // Test time
  \STATE $\qbup=round(\bup.q)$
  \STATE $psnr=\mathbf{d}(\mathbf{x},g_s(\mathbf{\hat{y}};\mathbf{\theta}_g,\beta_g+\qbup/q))$
  \STATE $s_{min}=min(\qbup), s_{max}=max(\qbup)$
  \STATE $\mu^*,\sigma^*=argmax_{\mu,\sigma}(\prod_i\frac{N(\qbup_i;\mu,\sigma)}{\sum_{z=s_{min}}^{s_{max}} N(z;\mu,\sigma)})$
  \STATE $sp=\sum_{z=s_{min}}^{s_{max}} N(z;\mu^*,\sigma^*)$
  \STATE $rate=-log(p(\qbup;\mu^*,\sigma^*)/sp)+C$
  \STATE $acc=\frac{|mb|+|sb|+rate}{\mathbf{R}(psnr)}$
  \IF{$acc<best$ }
    \STATE $best=acc$; $\bup^{(b)}=\bup$; $q^{(b)}=q$; $s_{min}^{(b)}=s_{min}$;
    \STATE $s_{max}^{(b)}=s_{max}$; $\mu^{(b)}=\mu^*$; $\sigma^{(b)}=\sigma^*$; 
  \ENDIF
  \ENDFOR
  \STATE $eb.write(q^{(b)})$; $eb.write(\mu^{(b)})$; //Encode params explicitly 
  \STATE $eb.write(\sigma^{(b)})$; $eb.write(s_{min}^{(b)})$; $eb.write(s_{max}^{(b)})$; 
  \STATE $\qbup=round(\bup^{(b)}.q^{(b)})$
  \STATE $sp=\sum_{z=s_{min}}^{s_{max}} N(z;\mu^{(b)},\sigma^{(b)})$
  \STATE $eb.write(\qbup,p(.;\mu^{(b)},\sigma^{(b)})/sp)$ //Encode updates w.r.t fitted gaussian
  
\end{algorithmic}
\end{algorithm}

\begin{algorithm}[tb]
  \caption{Proposed Decoding}
  \label{alg:decode}
  \footnotesize
\begin{algorithmic}
  \STATE {\bfseries Input:}
  Main latent bitstream $mb$, side latent bitstream $sb$ and extra bitstream $eb$.
  Trained baseline model components   $g_s(.;\mathbf{\theta}_g,\beta_g)$, $p_{f}(.|\Psi)$, $p_{h}(.|.;\Theta)$.
  \STATE {\bfseries Output:} Reconstruction $\mathbf{\hat{x}}$
 \vspace{0.05cm}
 \hrule 
 \vspace{0.05cm}
  \STATE $\mathbf{\hat{z}}=sb.read(p_{f}(.|\Psi))$ // decode side latent w.r.t factorized entropy
  \STATE $\mathbf{\hat{y}}=mb.read(p_{h}(.|\mathbf{\hat{z}};\Theta))$ // decode main latent w.r.t hyperprior entropy
  \STATE $q^{(b)}=eb.read()$; $\mu^{(b)}=eb.read()$; //decode params explicitly
  \STATE $\sigma^{(b)}=eb.read()$; $s_{min}^{(b)}=eb.read()$; $s_{max}^{(b)}=eb.read()$; 
  \STATE $sp=\sum_{z=s_{min}}^{s_{max}} N(z;\mu^{(b)},\sigma^{(b)})$
  \STATE $\qbup=eb.read(p(.;\mu^{(b)},\sigma^{(b)})/sp)$ //decode updates w.r.t fitted gaussian
  \STATE $\mathbf{\hat{x}}=g_s(\mathbf{\hat{y}};\mathbf{\theta}_g,\beta_g+\qbup/q^{(b)})$
\end{algorithmic}
\end{algorithm}

\section{Results}
\label{sec:results}

\begin{figure*}[tb]
\includegraphics[width=1.0\linewidth]{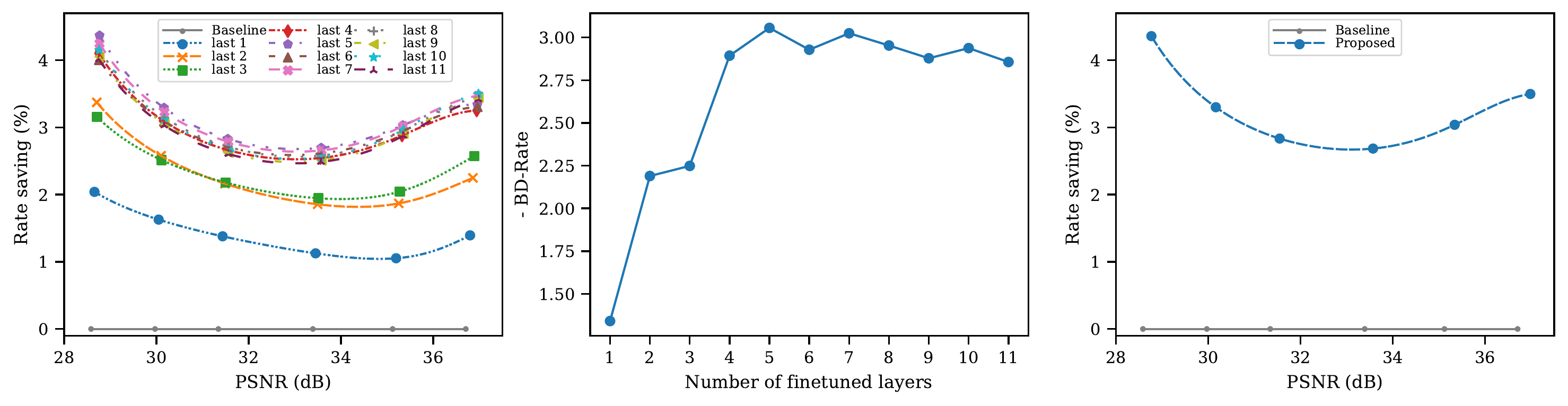}
\centering
\caption{Experimental results on Kodak Test Set for  \textbf{cheng2020-anchor} model in \cite{cheng2020image} trained on 6 different psnr objectives. The number of last layer's bias to be overfitted impacts the performance. The relative bit-savings varies with respect to the quality. In average we have $3\%$ gain.}
\label{fig:res_kodak_cheng}
\end{figure*}
\begin{figure*}[tb]
\includegraphics[width=1.0\linewidth]{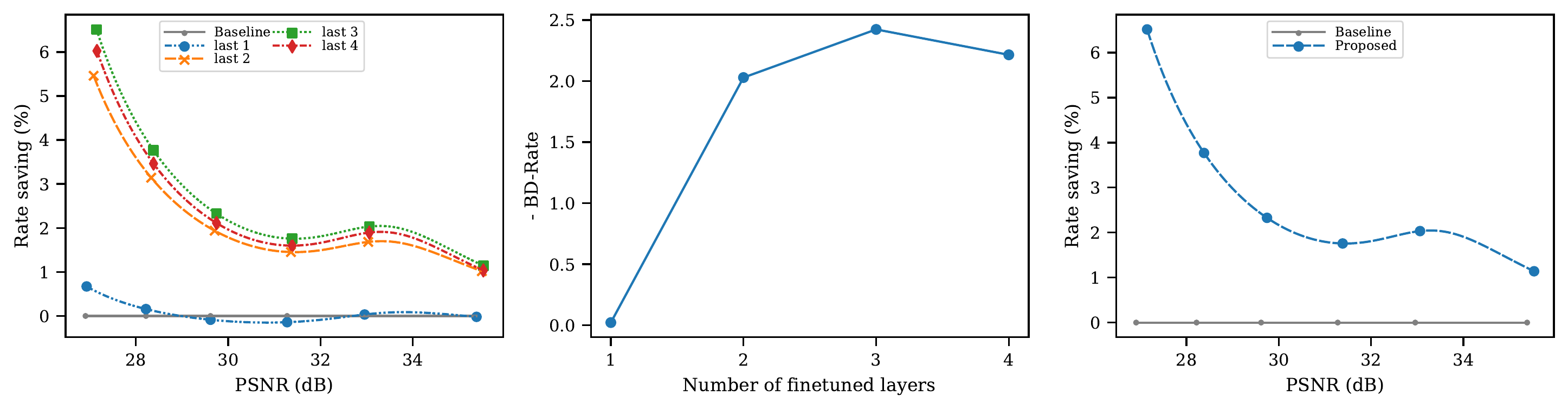}
\centering
\caption{Experimental results on Kodak Test Set for \textbf{bmshj2018-factorized} model in \cite{balle2018variational} trained on 6 different psnr objectives.}
\label{fig:res_kodak_balle}
\end{figure*}

\begin{figure*}[tb]
\includegraphics[width=1.0\linewidth]{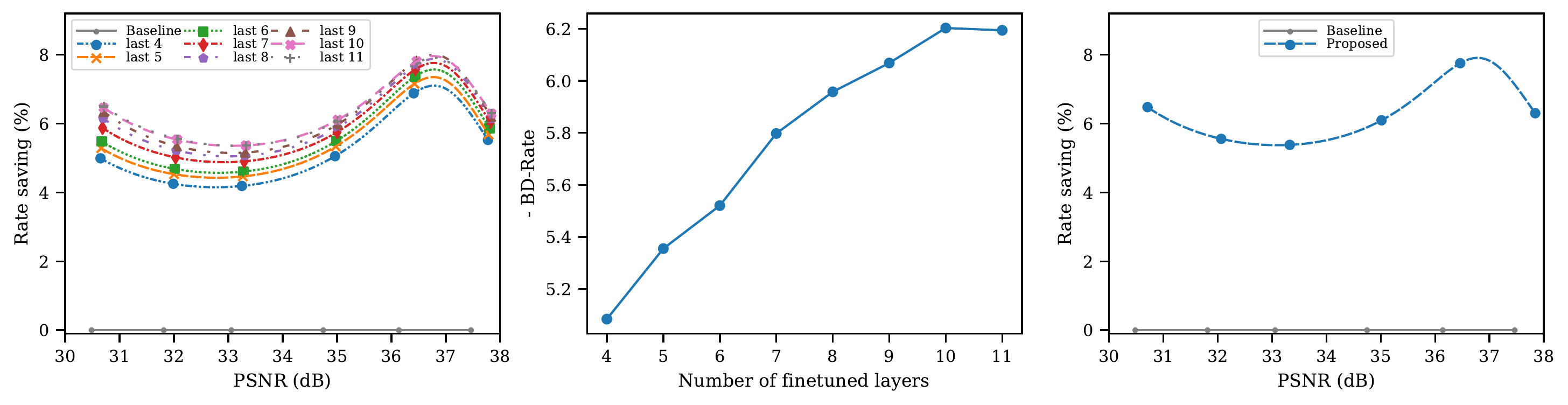}
\centering
\caption{Experimental results on CLIC Test Set for \textbf{cheng2020-anchor} model trained on 6 different psnr objectives.}
\label{fig:res_clic_cheng}
\end{figure*}

\begin{figure*}[tb]
\begin{center}
    \includegraphics[width=0.32\linewidth]{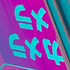}
    \includegraphics[width=0.32\linewidth]{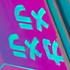}
    \includegraphics[width=0.32\linewidth]{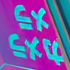}
\end{center}

\caption{Illustration of our approach on image 22 of the CLIC Test Set for \textbf{cheng2020-anchor}: the original (left), ours (middle) and the corresponding unimproved baseline (right). The figure shows only a patch of the image.}
\label{fig:illu}
\end{figure*}

In this section, we empirically show the strength of our approach. We use as a baseline \textbf{\cheng} (\textbf{cheng2020-anchor}\cite{cheng2020image}), one of the strongest neural image compression models, and the Kodak dataset \cite{eastman_kodak_kodak_nodate}. To show the generalization of our approach, we also perform experiments using the CLIC 2021 test dataset \cite{CLIC} and with a different baseline: \bmshj\  (\textbf{bmshj2018-factorized} \cite{balle2018variational}), an early neural compression model.
For both baselines, we used their implementation in CompressAI~\cite{compressai}.
Algorithm \ref{alg:encode} runs for any given single test image $\mathbf{x}$ and every single pre-trained model in CompressAI library for different evaluations on RD curve. 
CompressAI provides 6 pre-trained \cheng\ models that give different bit-rates. We limit the training to $2500$ iterations and use Adam optimizer with a learning rate of $1e-3$. We initialize $b$ with all zero (no update) and $q=10$. For $\mathbf{R}(\mathbf{D})$ function, we use a linear interpolation between two pre-trained models: the baseline model and the one closest to the baseline in terms of quality. Our approach improves the overall quality of the decoded image and more specifically in the high-frequency parts, such as edges or text. An example is given in Fig. \ref{fig:illu}

\textbf{Our approach leads to improvements in every configuration tested.} Fig \ref{fig:res_kodak_cheng} to \ref{fig:res_clic_cheng}, right column, show the rate savings with respect to the baseline as a function of PSNR, when fine-tuning the best number of layers for each model. Our approach reduces the bitlength by at least $3\%$ under the same reconstruction quality for \cheng\ on Kodak (Fig \ref{fig:res_kodak_cheng}), between 1 and $6\%$ for \bmshj\ on Kodak (Fig \ref{fig:res_kodak_balle}) and by more than $5\%$ for \cheng\ on CLIC (Fig \ref{fig:res_clic_cheng}). This suggests our approach achieves larger improvements for state-of-the-art models than older ones. This might be due to the typically larger number of parameters of more recent models, which could overfit the training data more.

\textbf{The optimal number of layers to overfit is not the same for every baseline model.}
The left and middle columns of  Fig \ref{fig:res_kodak_cheng} to \ref{fig:res_clic_cheng} illustrate the impact of the number of layers whose bias are fine-tuned. The left plots show the rate savings for different baseline models whereas the middle one provides the average (over baseline models) BD rate as a function of the number of layers fine-tuned.
 As expected, the performance is at first increasing in all cases until a given number of layers (respectively $5$, $3$ and $10$  in Fig \ref{fig:res_kodak_cheng}, \ref{fig:res_kodak_balle} and \ref{fig:res_clic_cheng}). Increasing the number of layers further tends to slightly decrease the gains: sending additional weight updates leads to comparatively smaller improvements in performance. Furthermore, the optimal number of layers is larger for CLIC (10) than for Kodak (5) and \cheng. This is due to the size of the images. As the images are larger in CLIC, the cost of updating more weights is relatively smaller compared to the encoding of the image. Furthermore, larger images typically contain more details, increasing potential gains.

\textbf{Last layers are the best candidate for overfitting.}
We experimented with various subsets of layers of the decoder. We tried out the biases in the first layers, last layers, and randomly selected layers. The results achieved when using the last layers consistently surpass the others. This is expected because the details of a picture are usually learnt by the last layers. Hence we only reported results for the last layers.

\section{Conclusion}
\label{sec:conc}
In this work, we improve the BD-rate of state-of-the-art neural image compression by 3-5\% by finetuning part of the decoder, thus reducing the amortization gap. 
Decoder finetuning was so far limited 
to video or set of images, as the cost of encoding the weight updates was too large for single images. Decoder finetuning is now feasible for single image compression thanks to our three contributions: the defined hyperparameter free new loss function that explicitly guaranties the model obtains better RD curve performance and avoids hyperparameter optimization; subset selection of decoder's parameters and learnable quantization scale parameter. Analyzing additional baseline models, extending this approach for neural video compression, and finding the best subset of parameters in end to end manner would be our future line of research.  

\IEEEtriggeratref{13}

\bibliographystyle{IEEEtran}
\bibliography{string}

\begin{thebibliography}{10}
\providecommand{\url}[1]{#1}
\csname url@samestyle\endcsname
\providecommand{\newblock}{\relax}
\providecommand{\bibinfo}[2]{#2}
\providecommand{\BIBentrySTDinterwordspacing}{\spaceskip=0pt\relax}
\providecommand{\BIBentryALTinterwordstretchfactor}{4}
\providecommand{\BIBentryALTinterwordspacing}{\spaceskip=\fontdimen2\font plus
\BIBentryALTinterwordstretchfactor\fontdimen3\font minus
  \fontdimen4\font\relax}
\providecommand{\BIBforeignlanguage}[2]{{%
\expandafter\ifx\csname l@#1\endcsname\relax
\typeout{** WARNING: IEEEtran.bst: No hyphenation pattern has been}%
\typeout{** loaded for the language `#1'. Using the pattern for}%
\typeout{** the default language instead.}%
\else
\language=\csname l@#1\endcsname
\fi
#2}}
\providecommand{\BIBdecl}{\relax}
\BIBdecl

\bibitem{vvc}
B.~Bross, J.~Chen, J.-R. Ohm, G.~J. Sullivan, and Y.-K. Wang, ``Developments in
  international video coding standardization after avc, with an overview of
  versatile video coding (vvc),'' \emph{Proceedings of the IEEE}, vol. 109,
  no.~9, pp. 1463--1493, 2021.

\bibitem{gao2021neural}
G.~Gao, P.~You, R.~Pan, S.~Han, Y.~Zhang, Y.~Dai, and H.~Lee, ``Neural image
  compression via attentional multi-scale back projection and frequency
  decomposition,'' in \emph{Proceedings of the IEEE/CVF International
  Conference on Computer Vision}, 2021, pp. 14\,677--14\,686.

\bibitem{habibian2019video}
A.~Habibian, T.~v. Rozendaal, J.~M. Tomczak, and T.~S. Cohen, ``Video
  compression with rate-distortion autoencoders,'' in \emph{Proceedings of the
  IEEE/CVF International Conference on Computer Vision}, 2019, pp. 7033--7042.

\bibitem{kingma2013auto}
D.~P. Kingma and M.~Welling, ``Auto-encoding variational bayes,'' in
  \emph{ICLR}, 2013.

\bibitem{balle2018variational}
J.~Ball{\'e}, D.~Minnen, S.~Singh, S.~J. Hwang, and N.~Johnston, ``Variational
  image compression with a scale hyperprior,'' in \emph{ICLR}, 2018.

\bibitem{1056282}
J.~Rissanen and G.~Langdon, ``Universal modeling and coding,'' \emph{IEEE
  Transactions on Information Theory}, vol.~27, no.~1, pp. 12--23, 1981.

\bibitem{theis2017lossy}
L.~Theis, W.~Shi, A.~Cunningham, and F.~Husz{\'a}r, ``Lossy image compression
  with compressive autoencoders,'' in \emph{ICLR}, 2017.

\bibitem{balle2016end}
J.~Ball{\'e}, V.~Laparra, and E.~P. Simoncelli, ``End-to-end optimized image
  compression,'' in \emph{ICLR}, 2017.

\bibitem{minen_joint}
D.~Minnen, J.~Ball\'{e}, and G.~D. Toderici, ``Joint autoregressive and
  hierarchical priors for learned image compression,'' in \emph{Advances in
  Neural Information Processing Systems}, S.~Bengio, H.~Wallach, H.~Larochelle,
  K.~Grauman, N.~Cesa-Bianchi, and R.~Garnett, Eds., vol.~31, 2018.

\bibitem{minnen2020channel}
D.~Minnen and S.~Singh, ``Channel-wise autoregressive entropy models for
  learned image compression,'' in \emph{ICIP}, 2020, pp. 3339--3343.

\bibitem{cheng2020image}
Z.~Cheng, H.~Sun, M.~Takeuchi, and J.~Katto, ``Learned image compression with
  discretized gaussian mixture likelihoods and attention modules,'' in
  \emph{CVPR}, 2020.

\bibitem{xie2021enhanced}
Y.~Xie, K.~L. Cheng, and Q.~Chen, ``Enhanced invertible encoding for learned
  image compression,'' in \emph{Proceedings of the ACM International Conference
  on Multimedia}, 2021.

\bibitem{cremer2018inference}
C.~Cremer, X.~Li, and D.~Duvenaud, ``Inference suboptimality in variational
  autoencoders,'' \emph{ICML}, 2018.

\bibitem{LuCZCOXG20}
G.~Lu, C.~Cai, X.~Zhang, L.~Chen, W.~Ouyang, D.~Xu, and Z.~Gao, ``Content
  adaptive and error propagation aware deep video compression,'' in
  \emph{ECCV}, vol. 12347, 2020, pp. 456--472.

\bibitem{Campos_2019_CVPR_Workshops}
J.~Campos, S.~Meierhans, A.~Djelouah, and C.~Schroers, ``Content adaptive
  optimization for neural image compression,'' in \emph{CVPR Workshops}, June
  2019.

\bibitem{NEURIPS2020_066f182b}
Y.~Yang, R.~Bamler, and S.~Mandt, ``Improving inference for neural image
  compression,'' in \emph{Advances in Neural Information Processing Systems},
  H.~Larochelle, M.~Ranzato, R.~Hadsell, M.~F. Balcan, and H.~Lin, Eds.,
  vol.~33, 2020, pp. 573--584.

\bibitem{Guo_2020_CVPR_Workshops}
T.~Guo, J.~Wang, Z.~Cui, Y.~Feng, Y.~Ge, and B.~Bai, ``Variable rate image
  compression with content adaptive optimization,'' in \emph{CVPR Workshops},
  June 2020.

\bibitem{9287069}
N.~Zou, H.~Zhang, F.~Cricri, H.~R. Tavakoli, J.~Lainema, M.~Hannuksela,
  E.~Aksu, and E.~Rahtu, ``L2c – learning to learn to compress,'' in
  \emph{2020 IEEE 22nd International Workshop on Multimedia Signal Processing
  (MMSP)}, 2020, pp. 1--6.

\bibitem{Aytekin_2018_CVPR_Workshops}
C.~Aytekin, X.~Ni, F.~Cricri, J.~Lainema, E.~Aksu, and M.~Hannuksela,
  ``Block-optimized variable bit rate neural image compression,'' in \emph{CVPR
  workshop}, June 2018.

\bibitem{van2021overfitting}
T.~van Rozendaal, I.~A. Huijben, and T.~S. Cohen, ``Overfitting for fun and
  profit: Instance-adaptive data compression,'' in \emph{ICLR}, 2021.

\bibitem{van2021instance}
T.~van Rozendaal, J.~Brehmer, Y.~Zhang, R.~Pourreza, and T.~S. Cohen,
  ``Instance-adaptive video compression: Improving neural codecs by training on
  the test set,'' \emph{arXiv preprint arXiv:2111.10302}, 2021.

\bibitem{bjontegaard2001calculation}
G.~Bjontegaard, ``Calculation of average psnr differences between rd-curves,''
  \emph{VCEG-M33}, 2001.

\bibitem{eastman_kodak_kodak_nodate}
\BIBentryALTinterwordspacing
E.~Kodak, ``Kodak {Lossless} {True} {Color} {Image} {Suite} ({PhotoCD}
  {PCD0992}).'' [Online]. Available: \url{http://r0k.us/graphics/kodak}
\BIBentrySTDinterwordspacing

\bibitem{CLIC}
``{CLIC2021}: Challenge on learned image compression,''
  \url{http://clic.compression.cc/2021/}
  year = {2021}.

\bibitem{compressai}
J.~B{\'e}gaint, F.~Racap{\'e}, S.~Feltman, and A.~Pushparaja, ``Compressai: a
  pytorch library and evaluation platform for end-to-end compression
  research,'' \emph{arXiv preprint arXiv:2011.03029}, 2020.

\end{thebibliography}


\end{document}